\documentclass[doublespacing]{elsart}
\usepackage{stmaryrd}
\usepackage{amsmath}
\usepackage{amssymb}
\usepackage{amsfonts}
\usepackage{graphicx}
\usepackage{rotating}
\usepackage{longtable}
\usepackage{graphics}
\usepackage{epstopdf}
\usepackage{epsfig}
\usepackage{color}

\date{}
\begin{document}

\noindent \textbf{Corresponding author: }\\
\noindent Prof. Dr. Hong-Jian Feng\\
\noindent
Department of Physics,\\
Northwest University,\\ Xi'an 710069, People's Republic of China\\
Tel.:
+86-29-88303384\\
Email address:\\
hjfeng@nwu.edu.cn\\
fenghongjian@gmail.com\\
\clearpage

\begin{frontmatter}

\title{Electric-field switching magnetization  and spin transfer   in  ultrathin BiFeO$_3$ film}

\author{Hong-Jian  Feng\corauthref{cor1}}\ead{hjfeng@nwu.edu.cn, fenghongjian@gmail.com}

\address{ Department of  Physics, Northwest University, Xi'an 710069, People's Republic of China}

\corauth[cor1]{Corresponding author.}



\begin{abstract}
First-principles density-functional theory calculations  show  switching  magnetization by 90\textordmasculine can be achieved in ultrathin BFO film by applying external electric-field. Up-spin carriers  appear to the surface with positive field while down-spin ones to the negative field surface, arising from the redistribution of Fe-t$_{2g}$ orbital. The half-metallic behavior of Fe-$3d$ states  in the surface of R phase film makes it a promising candidate for AFM/FM bilayer heterostructure possessing electric-field tunable FM magnetization reversal and opens a new way towards designing spintronic multiferroics. The interface exchange-bias effect in this BFO/FM bilayer is mainly driven by the Fe-t$_{2g}$ orbital reconstruction, as well as spin transferring and rearrangement.
\end{abstract}

\end{frontmatter}


\section{ Introduction}
BiFeO$_3$(BFO) attracted much attention due to the coexistence of  antiferromagnetic(AFM) and
ferroelectric(FE) behavior, as well as the enhanced weak ferromagnetic(FM) magnetization  and FE polarization in
films on SrTiO$_3$ substrates prepared by pulsed laser deposition(PLD)\cite{1}. The enhanced FM and FE properties in
BFO films are attributed to the suppression of cycloidal magnetic structure with a long
wavelength period of $\sim$ 620 {\AA}\cite{2} caused by the Dzyaloshinskii-Moriya
interaction(DMI) in bulk\cite{3,4}.  DMI is able to produce the canting of neighboring Fe ions with
G-type AFM  structure observed under a N\'{e}el temperature
of $\sim$ 643 K\cite{5}, along with the spiral magnetization along the crystal.  In terms of magnetization enhancement,
DMI gives rise to the tilting of  the neighboring G-type AFM Fe ions leading to the weak FM magnetization, and the side-effect is the
spiral spin  structure which tends to reduce the macroscopic magnetization. Fortunately, doping\cite{6,7,8} and
thin film fabrication\cite{1}can be used to impede the spiral magnetization owing to the suppression of magnetic cycloid by doping ions and size-effect in epitaxial films. However, from
a practical point of view  several problems have to be solved:(\expandafter{\romannumeral1}) The weak magnetoelectric coupling  is quadratic due to the spin cycloid in the single phase BFO\cite{9,10,11}. (\expandafter{\romannumeral2})The indirect magnetoelectric coupling
 is mediated by the direct coupling between  antiferrodistortive(AFD) distortions being the out of phase rotation of neighboring oxygen octahedra along [1 1 1] direction and AFM vectors\cite{12,13,14}.(\expandafter{\romannumeral3}) Electric-field controlled FM magnetization is not able to be implemented because of the FE-AFM coupling in single phase BFO. In order to drive FM magnetization reversal by external electric-field in
  magnetic memory storage with lower energy consumption, a heterostructure composed by the BFO AFM layer and FM layer,e.g. BFO/La$_{0.7}$Sr$_{0.3}$MnO(LSMO) heterostructure\cite{15,16},is proposed to accomplish it instead of using the single phase BFO films where the quadratic magnetoelectric coupling is FE-AFM interaction\cite{17,18}.
  Electric-field tunable FM magnetization in  BFO/FM heterostructure is mainly  achieved through exchange-bias(EB) effect in the heterointerface\cite{19,20,21},and the EB coupling effect is more than orbital reconstruction, the AFM magnetization rotation, spin transfer, and electronic structure  in the interface are all playing a role to it, so exploring the above-mentioned properties in BFO film would be a significant step towards understanding the origin of it.   Motivated by the skin layer recently observed in the BFO single crystals\cite{22},  we proposed an ultrathin BFO film with considering the rhombohedral $R3c$ (R)phase and the tetragonal $P4mm$(P)phase recently observed in the morphotropic phase boundary(MPB) in BFO film\cite{23}. We carried out density functional theory(DFT) based calculations to investigate the AFM magnetization  rotation and the microscopic mechanism involved under external electric-field.

\section{ Computational details}
We use the local spin density
approximation(LSDA) to DFT scheme with a uniform energy cutoff of 500 eV  as in our previous work
\cite{12,13,14,24}. We considered Bi 5s, 5p, and 6s electrons, Fe 3s, 3p, and 3d electrons, and O 2s and
2p electrons as valence states and use $11\times11\times1$ Monkhorst-Pack sampling of the Brillouin zone
 for all calculations. We used the slab model to  construct (0 0 1) surface in hexagonal frame of reference
for R and P phase film having same vacuum  region as the atomic layers. The relaxation was carried out
with the top and bottom three atomic layers being moved as the forces on the ions were less than 0.01 meV/\AA.
G-type AFM spin configuration was used to construct the initial AFM structure, and the angel between the AFM direction
and the [1 1 1] FE polarization direction in rhombohedral representation was changing from 0\textordmasculine
 to 180\textordmasculine   to calculate the magnetic anisotropic energy(MAE) and the favorable spin
 configuration  with considering the spin-orbit coupling(SOC) and noncollinear magnetism. In addition, considering the
 FM ordering with the same  condition as mentioned in AFM case, we set the same variation of angle between FM spin and
 [1 1 1] direction to calculate the MAE  in comparison with AFM structure. We did not introduce
  the on-site Coulomb interaction by adding a Hubbard-like term to the effective potential in that DFT+U scheme
   is susceptible to the DMI\cite{12}, and further the MAE.
  A saw-tooth like external electric potential\cite{25,26} has been used as
\begin{equation}
V_{ext}(\textbf{r})=4\pi m(\textbf{r}/r_m-1/2),  0<\textbf{r}<r_m
\end{equation}
where m is the surface dipole density of the slab, r$_m$ is the periodic length along the direction perpendicular to the slab. We used  electric-field of 2 V/${\AA}$ for R phase film and  1 V/${\AA}$ for T phase film due to the  restoring forces and overshoot in the iteration
 process as electric-field is beyond that value.

\section{ Results and discussion}

The initial structure for R and P phase films are taken from Ref. [1], then ions in the vicinity of the surface are all fully relaxed under field. The crystal structure of the R phase ultrathin film before and after applying field is shown in Fig.1. It is apparent that the Bi ions are displacing significantly along the field  comparing with the case in Fe, implying the Fe ions are confined in oxygen octahedra and  slightly affected by the field. Interestingly, with applying an electric-field, Fe ions in P phase films are actively participating in the tetragonal distortions owing to the epitaxial strain in-plane. AFM ordering is energetically favored for R phase films before and after exerting a field ,as well as for P phase films under electric-field, but FM ordering is more favorable in P phase films without external bias due to the unreleased strain in-plane.  Considering about the anisotropic superexchange interaction\cite{27} and single-ion anisotropy\cite{28}, the MAE per formula unit before and after applying external bias were reported for R and P phase in Fig. 2 and Fig.3, respectively. Sinusoidal waveform of MAE is established for AFM  spins except the P phase films with FM ordering where anomalies exist around the peak showing that the FM spins are not stable comparing with the AFM ones, and the gaussian fitting curves are demonstrated in Fig.4 for this case.  Without applying  bias, the R phase films exhibit an AFM spins parallel to the [1 1 1] polarized  direction in rhombohedral representation, then spins rotate to direction perpendicular to the [1 1 1] direction under field. Moreover, the FM spins parallel to the [1 1 1] direction is also rotating to the direction perpendicular to the [1 1 1] in  the P phase films with applying electric-field.  Eventually,  an electric-field driven (1 1 1) easy magnetization plane is established in these two phase films, implying that a rotation of magnetization by 90\textordmasculine is able to be induced by the bias along [1 1 1] direction, and this phenomenon is expected to be observed in the skin layer of BFO bulk. From Fig.4 the gaussian fitting curves indicate that a local minimum MAE is calculated at $\sim$ 60\textordmasculine
relative  to the [1 1 1] direction, and this FM ordering is a metastable sate and prone to be overcome by the electric-field. The electric-field tuned rotation of magnetization and spin rearrangement in the ultrathin BFO film is depicted in Fig.5. A heterostructure  composed by this ultrathin BFO film and thin FM film is expected to possess an electric-field controlled FM magnetization reversal through the interface EB coupling.

In order to demonstrate further the magnetization and spin transfer behavior across the films under external bias, we reported the magnetization averaged in-plane across the R phase and P phase films in Fig.6 and Fig.7, respectively. One can  see that the up-spin  and down-spin carriers are equally spaced across the films, and they are equal in magnitude and opposite in direction which is caused by the AFM spin configuration. The maximum spin density is about 0.2 $\mu_B$ in R phase films which is less than 0.35 $\mu_B$  in P phase. The relative large value of magnetization in P phase is attributed to the epitaxial strain in-plane, and the tetragonal displacement of Fe ions out-of-plane which tends to couple with the oxygen octahedra rotation to produce a much higher spin density. We define the direction of a surface as the normal pointing out from it and the sign of the field  as the cosine of the angle between the direction of the field and surface, then there must be a positive field and negative field in two sides of a film for a given field. From Fig.6 and Fig.7, it is worth mentioning that the up-spin carriers are always moving along the electric-field while the down-spin carriers are in the opposite direction. Eventually, the up-spin carriers appear in the surface where  positive field exists, leaving the down-spin ones in the other side of the film with negative field. When the BFO skin layer is combined with the FM layer to achieve the bilayer heterostructure, the up-spin carriers are accumulating in the surface with positive electric-field as the down-spin ones  aggregating in the opposite side. The spin carriers are coupling to the FM layer through EB involving the orbital reconstruction and the spin transferring behavior, hereby the spin carriers would be reversed by the alternating electric-field leading to the electric-field driven FM magnetization reversal.

To shed light on the surface spin transfer and intrinsic electronic property, the density of states(DOS) of triply degenerate Fe-t$_{2g}$ orbital in R and P phase film surface with negative field are illustrated in Fig.8 and Fig.9, respectively. In BFO the Fe-$3d$ are decomposed into triply degenerate t$_{2g}$  and doubly degenerate e$_{g}$  states due to the crystal field splitting, and the latter is hardly to be find in the surface showing that the spin carriers are mostly arising from the former.  It is clear that the occupied Fe-t$_{2g}$ states are changing from majority spins to minority ones, indicating that the down-spin carriers are accumulating in the surface with negative field , and the corresponding phenomenon is able to be confirmed in the surface with positive field. The accumulation and rearrangement of spin carriers is determined by the redistribution of Fe-t$_{2g}$ orbital occupation with exerting external bias.  Moreover, there exists a small amount of minority spins found both in R and P phase films without external bias,and the majority spins are still dominating in the film. On the contrary, all the occupying states are minority spins under electric-field, this partly reflects the accumulating behavior in the surface which is consistent with the spin density calculation shown in Fig.6 and Fig.7. It is worth pointing that there are finite DOS of up-spin and down-spin Fe-t$_{2g}$ states passing through the fermi level in R phase film before and after turning on the bias, clearly showing the half-metallic property in Fe-$3d$ states. The half-metallic behavior achieved in the ultrathin BFO film enables it  a potential candidate for spintronic applications. Without field the occupying states are coming from up-spins in R phase film while the up-spin states passing through fermi energy is able to provide a transferring way across the surface, and the transferring path is provided by the down-spin ones under field. Therefore we suggest the half-metallic property in R phase film constitutes the interface EB coupling mechanism, and the R phase ultrathin BFO film is proposed to be an ideal candidate for making electric-field tunable FM magnetization reversal devices and spintronics.

\section{ Conclusion}
DFT based first-principles calculations have been performed to study the magnetization rotation and spin density in ultrathin BFO film. A (1 1 1) easy magnetization plane is achieved for R and P phase films with applying external bias, hereby an AFM magnetization rotation driven by electric-field is accomplished. The up-spin carriers are along the field direction and accumulating in the surface with positive field while the down-spin ones are opposing the field and aggregating on the other side of the film.  The half-metallic Fe-$3d$ states make R phase BFO ultrathin film  a prosperous candidate for spintronics.

\noindent\textbf{Acknowledgment}

This work was financially supported by the National Natural Science Foundation of China(NSFC) under Grant No.11247230(H.-J. F.),
and by the Science Foundation of Northwest University(Grant No.12NW12)(H.-J. F.)


\begin{thebibliography}{99}
{

\bibitem{1}
J. Wang, J. B. Neaton, H. Zheng, V. Nagarajan, S. B. Ogale, B. Liu,
D. Viehland, V. Vaithyanathan, D. G. Schlom, U. V. Waghmare, N. A.
Spaldin, K. M. Rabe, M. Wuttig, and R. Ramesh, Science 299,1719
(2003).
\bibitem{2}
I. Sosnowskat,T. Peterlin-Neumaier, and  E. Steichele,    J. Phys.
C: Solid State Phys. 15, 4835 (1982)
\bibitem{3}
P. W. Anderson, Phys. Rev. 115, 2 (1959).
\bibitem{4}
T. Moriya, Phys. Rev. Lett. 4, 228 (1960).
\bibitem{5}
G. Catalan and J. F. Scott, Adv. Mater. 21, 2463 (2009)
\bibitem{6}
N. Jeon, D. Rout, III. W. Kim, S.-J. L. Kang,   Appl. Phys. Lett., 98,072901 (2011)
\bibitem{7}
G. Catalan, K. Sardar, N.  Church,J.  Scott,R.  Harrison,S.  Redfern, Phys. Rev. B, 79,212415 (2009)
\bibitem{8}
Y. F. Cui, Y. G. Zhao, L. B.  Luo, J. J. Yang, H. Chang, M. H. Zhu, D. Xie, T. L. Ren,  Appl. Phys. Lett., 97,222904 (2010)
\bibitem{9}
A. K. Zvezdin, A. M. Kadomtseva, S. S. Krotov, A. P. Pyatakov, Y. F. Popov, G. P. Vorob'ev, J.  Mag. Mag. Mater. 300   224 (2006)
\bibitem{10}
 A. M. Kadomtseva, A. K. Zvezdin, Y. F. Popov , A. P. Pyatakov, G. P. Vorob'ev£¬ JETP Lett. 79, 571 (2004)
\bibitem{11}
I. Sosnowska, A. K. Zvezdin, J. Mag. Mag. Mater. 140, 167  (1995)
\bibitem{12}
H.-J. Feng,  J. Mag. Mag. Mater. 322   1765 (2010)
\bibitem{13}
H.-J. Feng,   J. Mag. Mag. Mater. 322  3755 (2010)
\bibitem{14}
H.-J. Feng,   J. Mag. Mag. Mater. 324  178 (2012)
\bibitem{15}
S.M. Wu, Shane A. Cybart, P. Yu, M. D. Rossell, J.X. Zhang, R. Ramesh and R. C. Dynes, Nature Mater. 9, 756 (2010)
\bibitem{16}
A.Y. Borisevich, H. J. Chang, M. Huijben, M.P. Oxley, S. Okamoto, M.K. Niranjan, J.D. Burton, E.Y. Tsymbal, Y. H. Chu, P. Yu, R. Ramesh, S.V. Kalinin, and S.J. Pennycook, Phys. Rev. Lett. 105, 087204(2010)
\bibitem{17}
R. Ramesh and N. A. Spaldin, Nature Mater. 6,21 (2007)
\bibitem{18}
S.W. Cheong and M. Mostovoy, Nature Mater. 6,13 (2007)
\bibitem{19}
J. Nogues, I. K. Schuller, J. Mag. Mag. Mater. 192,  203 (1999)
\bibitem{20}
D. Lebeugle, A. Mougin, M. Viret,D. Colson, J. Allibe, H. Bea, E. Jacquet, C. Deranlot, M. Bibes, A. Barthelemy, Phys. Rev. B , 81,13411 (2010)
\bibitem{21}
K. L. Livesey, Phys. Rev. B, 82, 064408, (2010)

\bibitem{22}
X. Marti,P. Ferrer, J. Herrero-Albillos, J. Narvaez,  V. Holy,  N. Barrett, M. Alexe, G. Catalan, Phys. Rev. Lett.   106, 236101 (2011)
\bibitem{23}
R. J. Zeches, M. D. Rossell, J. X. Zhang, A. J. Hatt, Q. He, C.-H. Yang, A. Kumar, C. H. Wang, A. Melville, C. Adamo, G. Sheng, Y.-H. Chu, J. F. Ihlefeld, R. Erni, C. Ederer, V. Gopalan, L. Q. Chen, D. G. Schlom, N. A. Spaldin, L. W. Martin, R. Ramesh, Science 326, 977 (2009)

\bibitem{24}
H.-J. Feng,  Physica B  DOI:10.1016/j.physb.2012.12.018  (2013)

\bibitem{25}
L. Bengtsson, Phys. Rev. B 59,12301 (1999)
\bibitem{26}
B. Meyer and David Vanderbilt, Phys. Rev. B 63, 205426 (2001)
\bibitem{27}
J. H. V. Vleck, Phys. Rev. 52, 1178 (1937)
\bibitem{28}
W. P. Wolf, Phys. Rev. 108, 1152 (1957)




 }
\end{thebibliography}


\clearpage \raggedright \textbf{Figure Captions:}

Fig.1 The structure of the R phase BFO ultrathin film before and after exerting field.

Fig.2  Dependence of total energy on the angle between spins and [1 1 1] polarized direction in R phase film with and without applying external bias, respectively.

Fig.3  Dependence of total energy on the angle between spins and [1 1 1] polarized direction in P phase film with and without applying external bias, respectively

Fig.4  Gaussian fitting of MAE in P phase film  without applying electric-field.

Fig.5 Schematic diagram for the magnetization rotation under electric-field, the dotted line denote the [1 1 1] direction, and the field is applying along the [1 1 1] direction.

Fig.6 Averaged spin density  in-plane  across the R phase ultrathin BFO film under external bias.

Fig.7 Averaged spin density  in-plane  across the P phase ultrathin BFO film under external bias.

Fig.8 t$_{2g}$ DOS for R phase film before and after applying electric-field.

Fig.9 t$_{2g}$ DOS for P phase film before and after applying electric-field.

\clearpage

 \begin{figure}
\centering
\includegraphics[width=11cm]{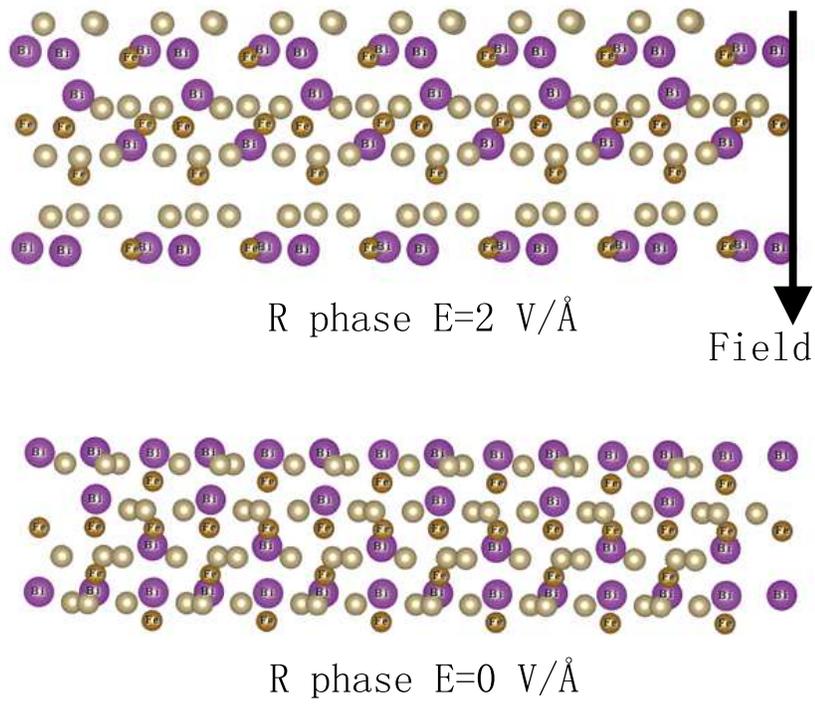}
\caption{{ The structure of the R phase BFO ultrathin film before and after exerting field.}}
\end{figure}

 \begin{figure}
\centering
\includegraphics[width=11cm]{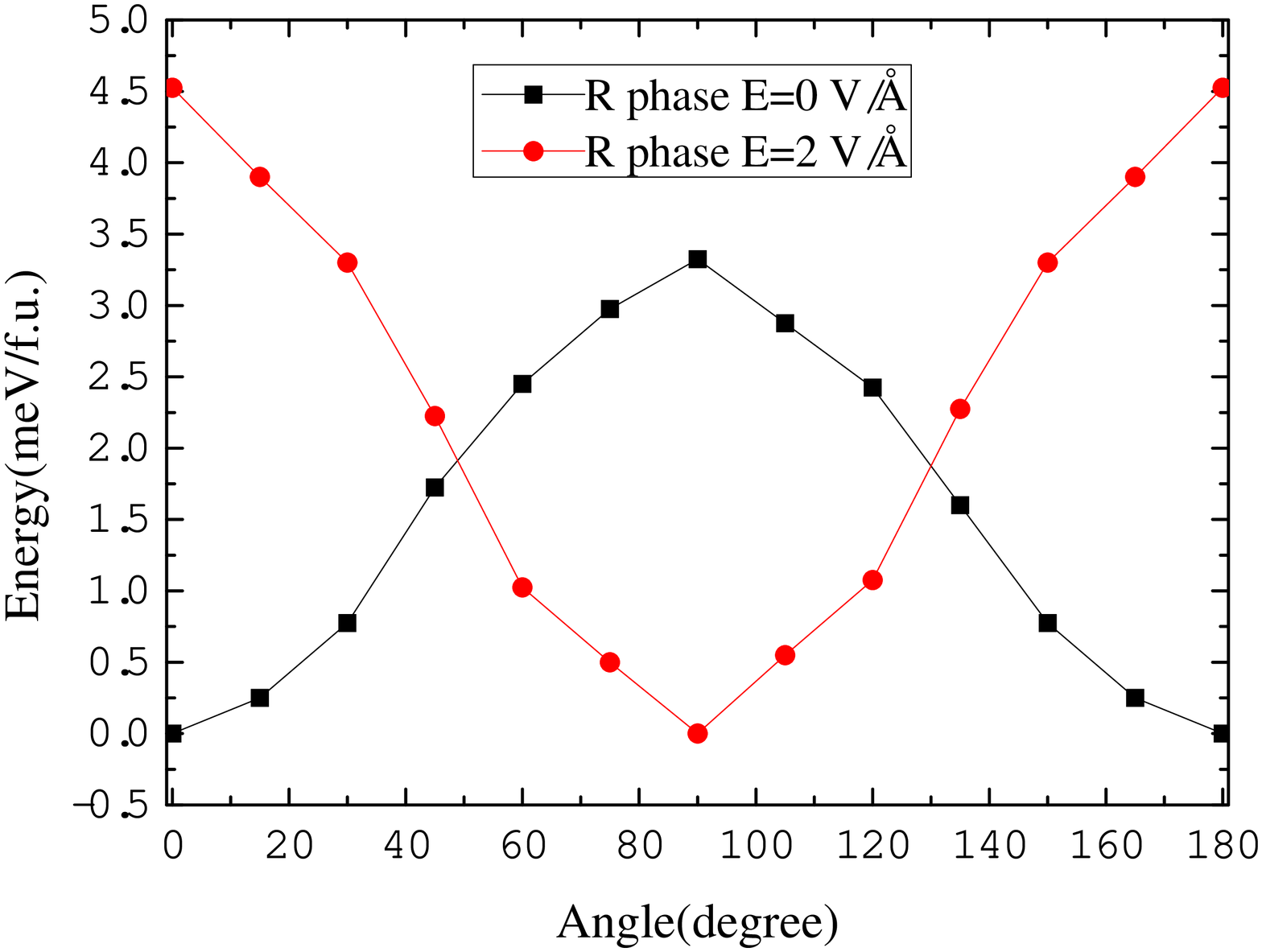}
\caption{{ Dependence of total energy on the angle between spins and [1 1 1] polarized direction in R phase film with and without applying external bias, respectively.}}
\end{figure}

 \begin{figure}
\centering
\includegraphics[width=11cm]{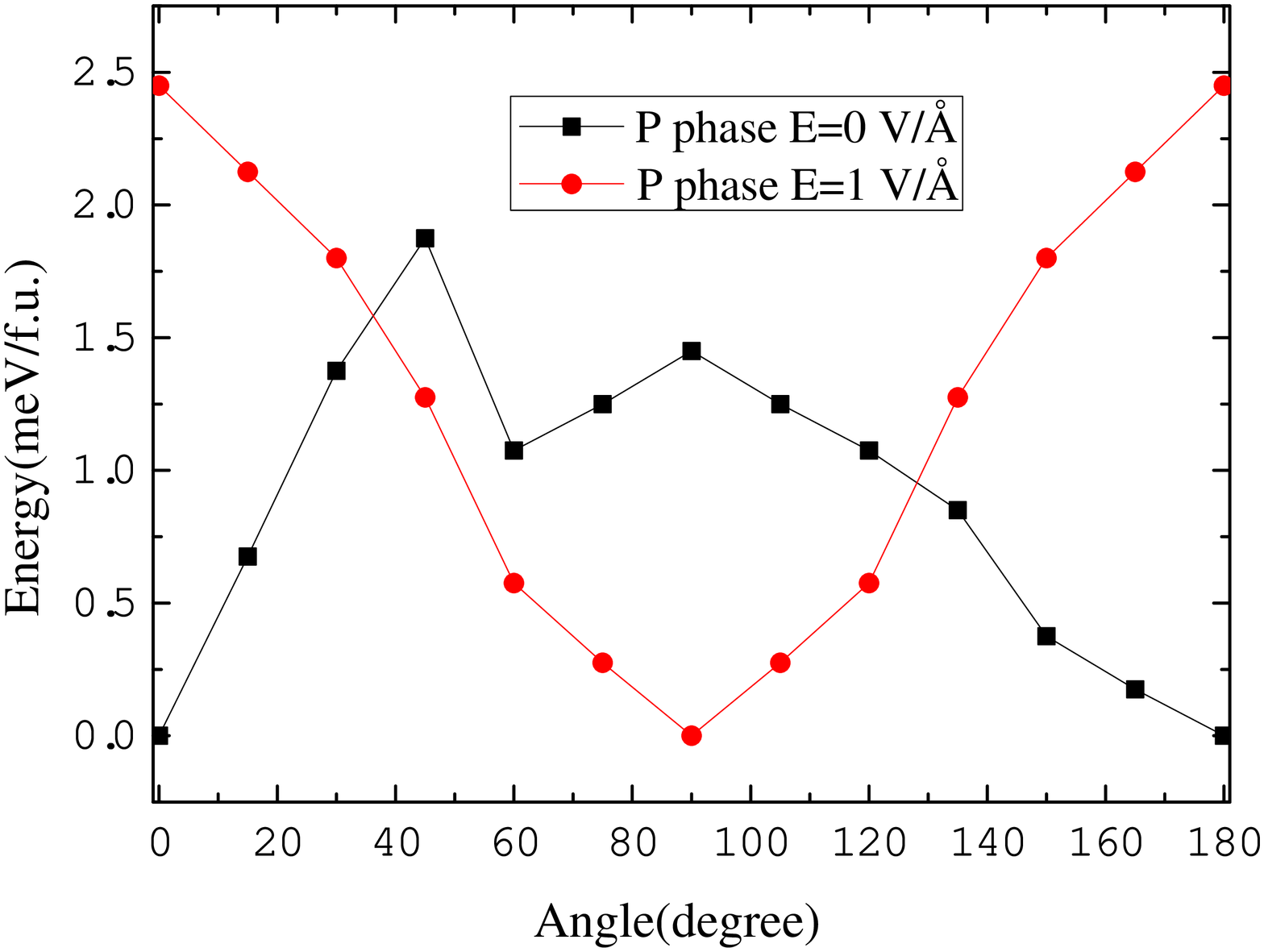}
\caption{{ Dependence of total energy on the angle between spins and [1 1 1] polarized direction in P phase film with and without applying external bias, respectively.}}
\end{figure}

 \begin{figure}
\centering
\includegraphics[width=11cm]{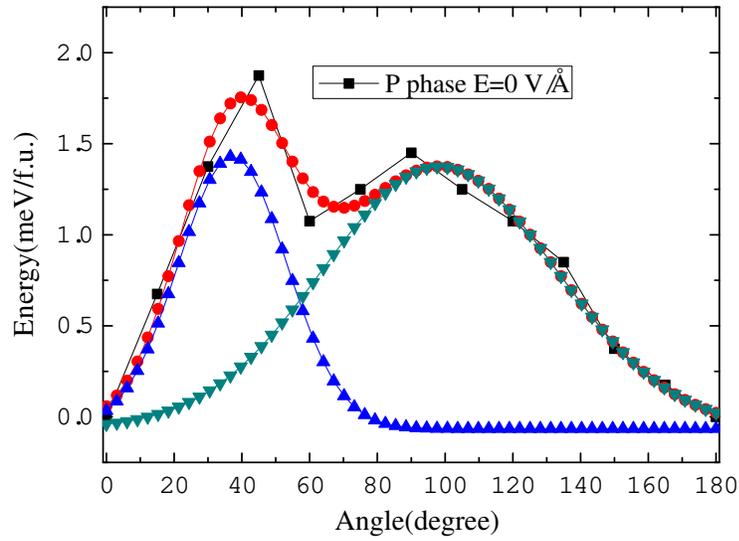}
\caption{{Gaussian fitting of MAE in P phase film  without applying electric-field.}}
\end{figure}

 \begin{figure}
\centering
\includegraphics[width=11cm]{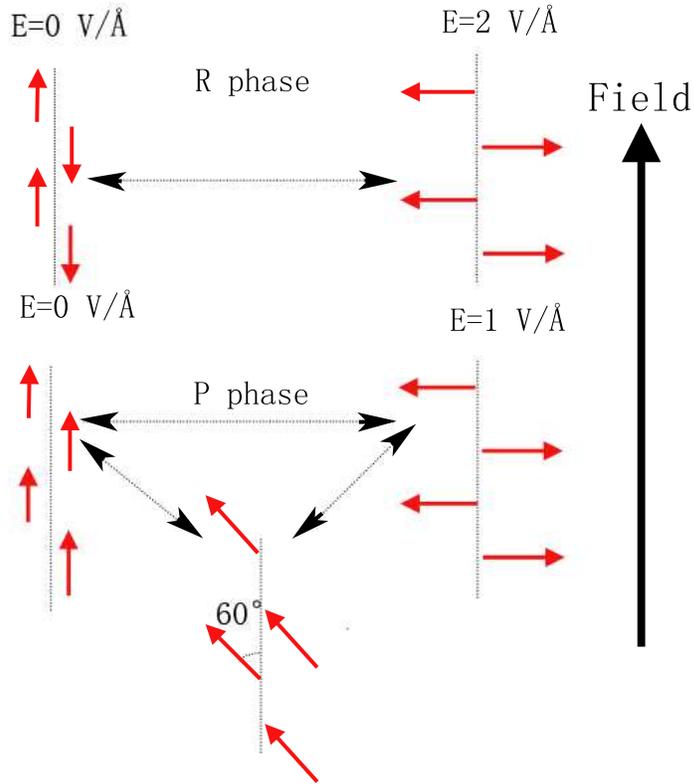}
\caption{{Schematic diagram for the magnetization rotation under electric-field, the dotted line denote the [1 1 1] direction, and the field is applying along the [1 1 1] direction.}}
\end{figure}

 \begin{figure}
\centering
\includegraphics[width=11cm]{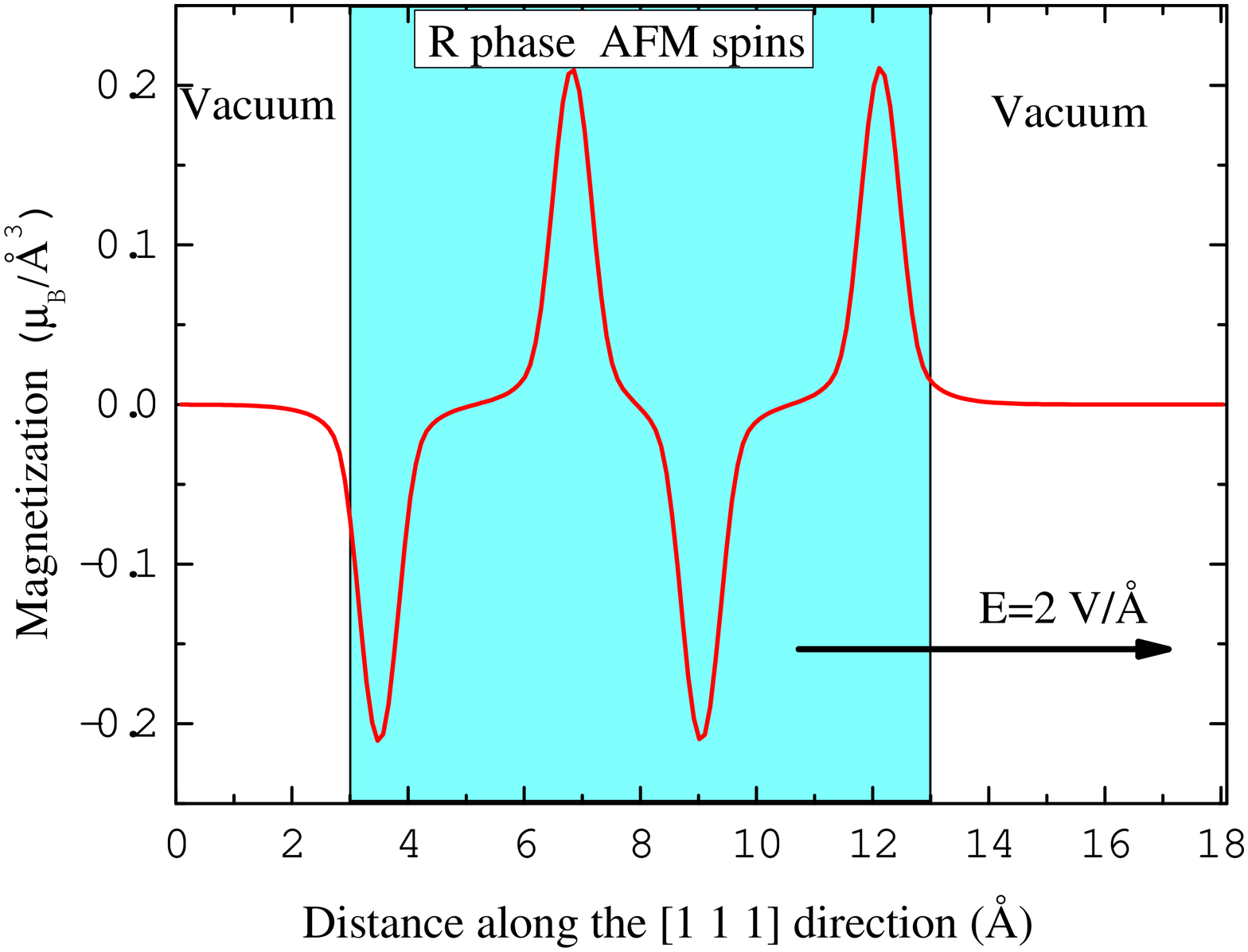}
\caption{{Averaged spin density  in-plane  across the R phase ultrathin BFO film under external bias.}}
\end{figure}

 \begin{figure}
\centering
\includegraphics[width=11cm]{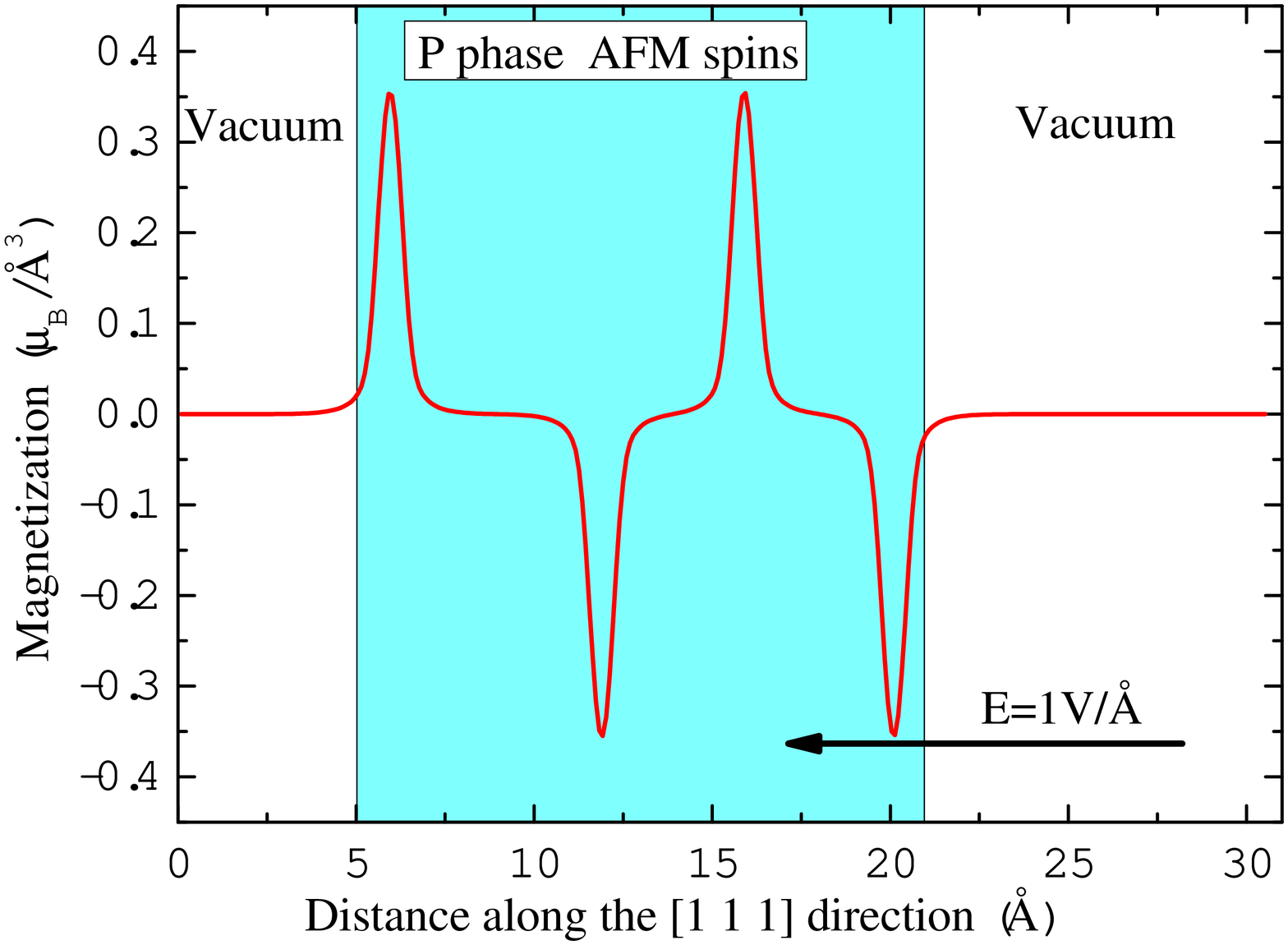}
\caption{{Averaged spin density  in-plane  across the P phase ultrathin BFO film under external bias.}}
\end{figure}

 \begin{figure}
\centering
\includegraphics[width=11cm]{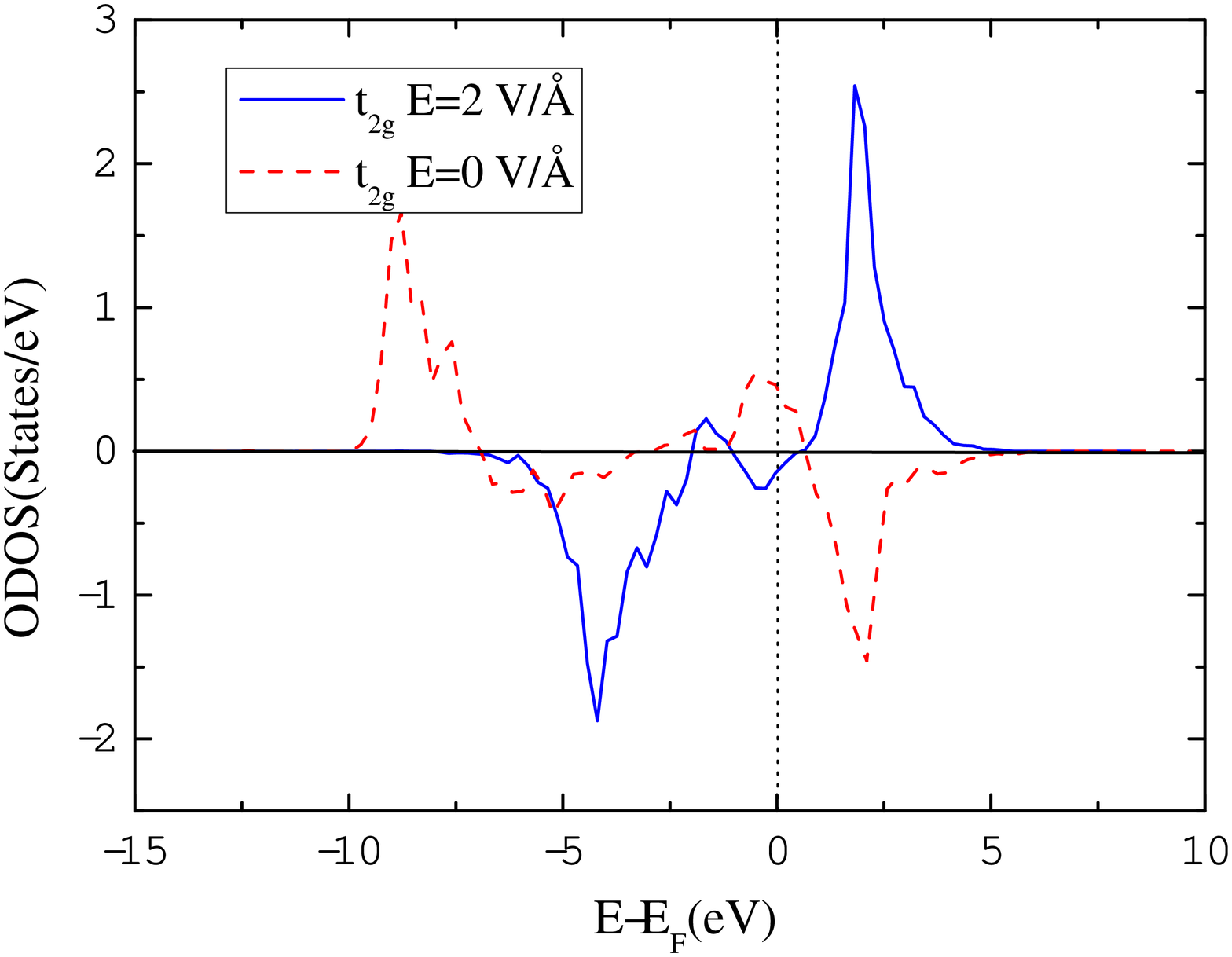}
\caption{{t$_{2g}$ DOS for R phase film before and after applying electric-field.}}
\end{figure}

 \begin{figure}
\centering
\includegraphics[width=11cm]{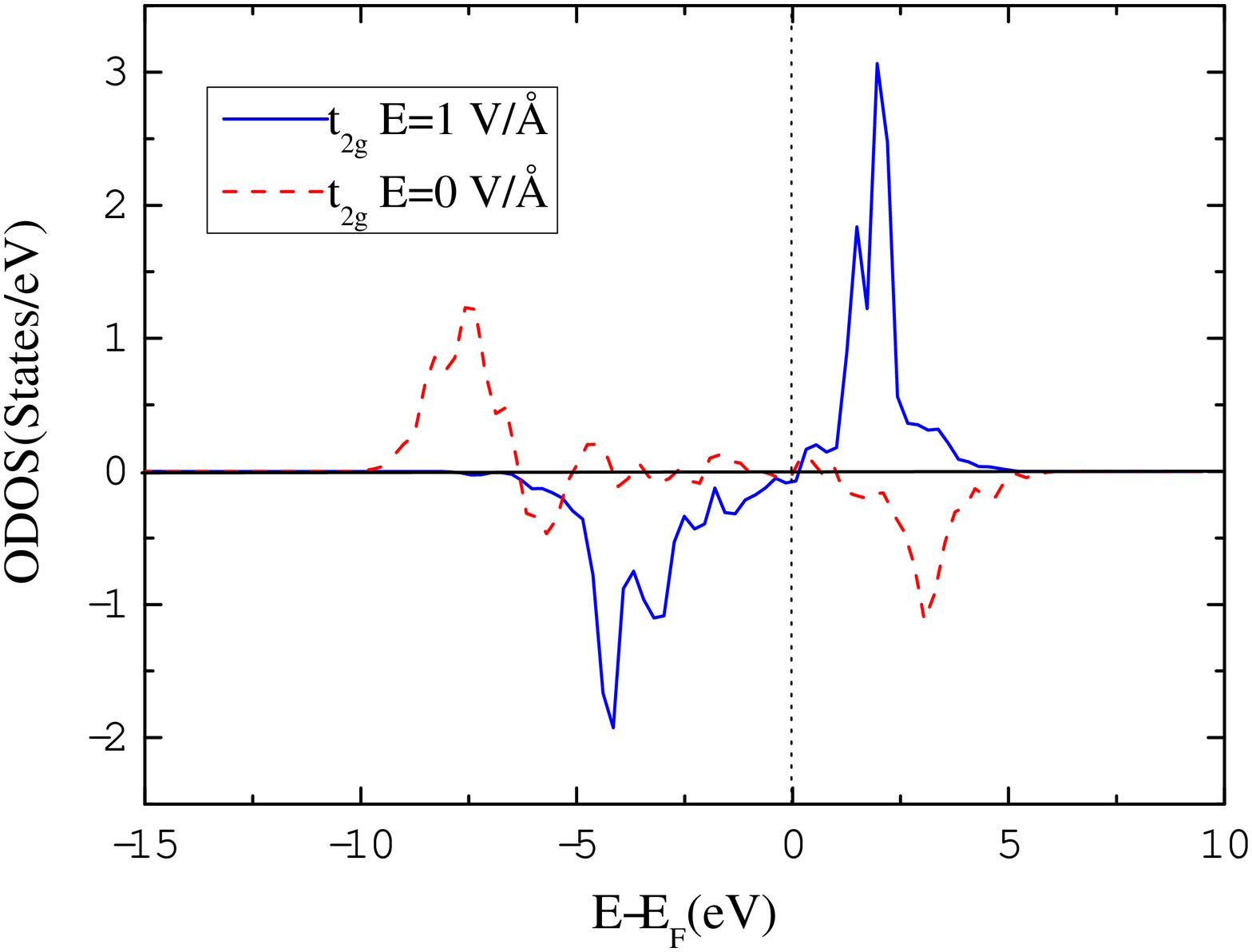}
\caption{{t$_{2g}$ DOS for P phase film before and after applying electric-field.}}
\end{figure}

\end{document}